\documentclass[apj]{emulateapj}
\usepackage{amsmath}
\usepackage{amssymb}
\usepackage{graphicx}
\usepackage{bm}
\DeclareMathOperator{\sech}{sech}

\begin{document}


\title{THE EFFECTS OF PLASMA BETA AND ANISOTROPY INSTABILITIES ON
THE DYNAMICS OF RECONNECTING MAGNETIC FIELDS IN THE HELIOSHEATH}

\author{K. M. Schoeffler} \author{J. F. Drake} \author{M. Swisdak}
\affil{Institute for Research in Electronics and Applied
Physics, University of Maryland, College Park, MD 20742-3511, USA}

\date{\today}

\begin{abstract}

The plasma $\beta$ (the ratio of the plasma pressure to the magnetic
pressure) of a system can have a large effect on its dynamics as high
$\beta$ enhances the effects of pressure anisotropies. We investigate the
effects of $\beta$ in a system of stacked current sheets that break up into
magnetic islands due to magnetic reconnection, which is analogous to the
compressed heliospheric current sheet in the heliosheath. We find significant
differences between systems with low and high initial values of $\beta$. At low
$\beta$ growing magnetic islands are modestly elongated and become round as
contraction releases magnetic stress and reduces magnetic energy. At high
$\beta$ the increase of the parallel pressure in contracting islands causes
saturation of modestly elongated islands as island cores approach the marginal
firehose condition. Only highly elongated islands reach finite size. The
anisotropy within these islands prevents full contraction, leading to a final
state of highly elongated islands in which further reconnection is suppressed.
The elongation of islands at finite $\beta$ is further enhanced by reducing the
electron-to-ion mass ratio, to more realistic values. The results are directly
relevant to reconnection in the sectored region of the heliosheath where there
is evidence that elongated islands are present, and possibly to other high
$\beta$ systems such as astrophysical accretion flows and the magnetosphere of
Saturn.

\end{abstract}

\maketitle

\section{Introduction}

At the outer edges of the solar system, the solar wind pushes up
against the interstellar medium. Since the solar wind is moving at
supersonic speeds, it forms a shock known as the termination shock.
The region between the termination shock and the interstellar medium
is referred to as the heliosheath. Flapping of the heliospheric
current sheet produces sectored magnetic fields \citep{Smith01} that have 
a significant latitudinal extent. There has been research suggesting that
the current sheets between the sectored fields found in the heliosheath are
compressed to the point that collisionless reconnection begins to occur,
resulting in the formation of magnetic islands \citep{Drake10,Czechowski10}. A
turbulent magnetohydrodynamic (MHD) model of the reconnection of the sectored
fields has also been proposed \citep{Lazarian09} although we will argue later
that the \emph{Voyager} data are inconsistent with this hypothesis.

An important question, however, is whether the conventional treatment of
collisionless reconnection \citep{Shay07} is valid in the heliosheath, where
it was suggested that the pick-up ion (PUI) population increases the plasma
pressure compared with values at 1 AU \citep{Zank99, Richardson08,Wu09}.
Although both \emph{Voyager} spacecraft are currently taking data in the heliosheath,
the energy range of the detectors does not cover the PUIs, so it is difficult
to make a reliable estimate of the value for $\beta$, the ratio of the plasma
pressure to the magnetic pressure \citep{Richardson08}. Global
MHD simulations suggest, however, that $\beta$ varies
from $8$ to $0.5$ between the termination shock and the interstellar medium
with the highest $\beta$ just downstream of the termination shock \citep{Drake10}.
Although this simulation does not include a separate pick-up ion population, it
provides a rough estimate for the expected values for $\beta$ and motivates
the range of $\beta$ in our study. In this study we investigate the impact of
$\beta$ on the dynamics of reconnection and the formation of magnetic islands
relevant to the sectored heliosheath.

To begin reconnection and island formation a current sheet needs to be
compressed to approximately the ion inertial scale $d_i = c/\omega_{\text{pi}}$, where
$c$ is the speed of light and $\omega_{\text{pi}}$ is the ion plasma frequency,
\citep{Cassak05,Yamada07}. At this point the current sheet becomes unstable to
the collisionless tearing mode. Upstream of the termination shock the
heliospheric current sheet has a thickness of around $10,000\text{ km}$
\citep{Smith01} and is predicted to compress to around $2500\text{ km}$ just
downstream of the shock. In the upstream region the plasma density measured at
\emph{Voyager 2} is around $0.001\text{cm}^{-3}$, corresponding to an ion skin depth of
around $7,200\text{ km}$.  In the downstream region the density is compressed
to around $0.003\text{cm}^{-3}$, corresponding to an ion skin depth of around
$4200\text{ km}$. Thus, the compression of the current sheets across the
termination shock should trigger collisionless reconnection in the heliosheath
and some \emph{Voyager 1} and \emph{2} observations support this hypothesis \citep{Opher11}. 

In our system we will be examining symmetric current sheets with no guide field
(initial out-of-plane magnetic field). In an analytic study
\cite{Brittnacher95} showed that when $\rho_i/w_0 \approx 1$, the fastest
growing linear mode occurs at $kw_0 \approx 0.5$, where $k$ is the wavenumber
of the tearing mode, $w_0$ is the half width of the current sheet, and $\rho_i$
is the ion gyroradius.  Since $\rho_i=\sqrt{\beta_i}d_i$, where $\beta_i$ is
the plasma beta based on the ion pressure, the current sheet thickness is
comparable to that in our simulation.

Due to the collisionless nature of the plasma, pressure anisotropies
($P_\parallel \neq P_\perp$, where $\parallel$ and $\perp$ are determined with
respect to the magnetic field) can form during reconnection. Fermi acceleration
in contracting islands \citep{Drake06a} and adiabatic cooling as $B$ decreases
due to conservation of the magnetic moment $\mu \propto v_\perp^2/B \propto
P_\perp/B$ both drive $P_\parallel > P_\perp$. When an anisotropy is formed
with $P_\parallel > P_\perp$ the tension in bent magnetic fields weakens. The
fluid momentum equation with an anisotropy becomes
\begin{equation}
\rho \frac{d\mathbf{v}}{dt} = -\mathbf{\nabla}\left(P_\perp + \frac{1}{8\pi} B^2 \right)
+ \mathbf{\nabla}\cdot\left[\left(1 - \frac{\beta_\parallel-\beta_\perp}{2}\right)
\frac{\mathbf{B}\mathbf{B}}{4 \pi} \right] \text{,}
\end{equation}
where $\rho$ is the mass density of the plasma, $\mathbf{v}$ is the bulk
velocity, and $\mathbf{B}$ is the magnetic field.
For $\beta_\parallel = \beta_\perp$ the pressure equation reduces to
the standard MHD equation. When
$\beta_\parallel > \beta_\perp$ the tension force is reduced. At
high $\beta$ this reduction of tension force is noticeable for even
slight anisotropies in the pressure. For $\beta_\parallel-\beta_\perp$
large enough, the tension force drops to zero, or even becomes
negative. Since the tension of field lines acts as a restoring
force for Alfv\'en waves in standard MHD, the negative sign causes
this oscillation to become an instability known as the firehose
instability \citep{Parker58} for:
\begin{equation}
  \label{firehose}
  \beta_\parallel - \beta_\perp > 2 \text{.}
\end{equation}
This instability is fueled by the free energy contained
in the pressure anisotropy. The firehose instability causes magnetic
field lines to kink, which eventually relieves the pressure anisotropy
by causing scattering.

Alternatively, when $\beta_\parallel-\beta_\perp$ is negative and
large enough in magnitude, other instabilities can occur. The mirror-mode instability and the ion cyclotron instability both occur when
$P_\perp > P_\parallel$. For larger $\beta_\parallel$ the mirror-mode
becomes unstable at smaller values of $|\beta_\parallel-\beta_\perp|$
than the ion cyclotron mode, so the marginal mirror-mode criterion
acts as the boundary between the stable and unstable regions. Based on
fluid theory assuming $T_e = T_i$ \citep{Hasegawa69}, the mirror-mode
instability occurs when
\begin{equation}
  \label{mirrormode}
  \beta_\perp - \beta_\parallel >
  \frac{\beta_\parallel}{\beta_\perp} \text{.}
\end{equation}

There are also kinetic modifications that can be made to the marginal
instability criteria for firehose, mirror-mode, and ion cyclotron
which make them more accurate. Although
a rigorous analytic theory is not available, there are models that
approximate the instability very well \citep{Hellinger06, Bale09}.
However, for simplicity we will just consider the conditions based
on fluid theory.  

In this study, we simulate several stacked current sheets similar to the
compressed sectored heliospheric fields and associated current sheets, and
follow the development of reconnection and islands.  We implement this system
in a two-dimensional particle-in-cell (PIC) code and vary the temperature of the background
plasma to test the dependence on $\beta$. We observe that in finite $\beta_e$
systems ($\beta_e > 0.5$), very elongated islands form as opposed to the modest-aspect-ratio islands found at low $\beta_e$ ($\beta_e < 0.5$), where $\beta_e$
is the $\beta$ based on the electron pressure. At high $\beta$, the increased
$P_\parallel$ due to the Fermi reflection of electrons within islands saturates
the normal modest-aspect-ratio islands. Fermi reflection in highly elongated
islands is less efficient because of the increased bounce time of the electrons
so these islands are able to reach finite amplitude. At late time, however,
even these elongated islands exhibit anisotropy instabilities, from Fermi
reflection of both ions and electrons. As a result, late-time magnetic islands
remain highly elongated and do not become round as in the low-$\beta$ regime.
This result has significant implications for the structure of islands that
would be measured in the heliosheath. Although $\beta_e$ is rather moderate in
the heliosheath, we find a mass ratio dependence suggesting long islands for a
broad range of $\beta_e$ in realistic mass ratios. A large $\beta$ however may
be necessary to sustain the elongation of these islands.

\section{Computational Model}

Our simulations are performed with the PIC code p3d. The initial
conditions consist of eight Harris current sheets \citep{harris62} where the
magnetic pressure balances the plasma pressure. Each Harris sheet consists of a
$\tanh(y/w_0)$ and $\sech^2(y/w_0)$ profile along the $\mathbf{\hat{y}}$
direction for an $\mathbf{\hat{x}}$-directed magnetic field and the density,
respectively. The peak density of the Harris sheets is $n_0$.  In addition,
there is a uniform background population that has a density of $n_b = 0.2n_0$.
These simulations are done in two dimensions so $\partial/\partial z = 0$, where
$\mathbf{\hat{z}}$ is out-of-plane, parallel to the initial current.

The code uses normalized units. The timescale is normalized to the
ion cyclotron time $\Omega_{\text{ci}}^{-1}$. The distance scales are
normalized to the ion inertial length $d_i = c/\omega_{\text{pi}}$. Thus, the
velocity is normalized to the Alfv\'en speed $v_A$. The magnetic field
is normalized to the asymptotic value of the reversed magnetic field
$B_0$. The density is normalized to $n_0$. The pressure is normalized
to $P_0 = n_0m_iv_A^2 = B_0^2/4\pi$. The temperature is normalized to
$T_0 = m_iv_A^2$.

In order to vary the $\beta$ of these simulations we vary the temperature of
the background population $T_b$. This background temperature is the same for
both ions and electrons. The Harris equilibrium is used to balance the sharp
change in the magnetic field strength across the current sheets, while the
background represents the PUIs and has the greatest influence on late-time reconnection dynamics.  We performed simulations for $\beta = 0.2$, $1$,
$2$, $3$, and $4.8$, where $\beta$ is based on the pressure in the asymptotic
field with density $n_b$.  Each simulation was advanced for a time of
$120\Omega_{\text{ci}}^{-1}$ with a time resolution $dt = 0.004\Omega_{\text{ci}}^{-1}$. The
simulations are on a $204.8d_i\times 102.4d_i$ domain with a grid scale
resolution of $\Delta_x = \Delta_y = 0.05d_i$. In order to complete such large
runs, unless otherwise specified, we used $25$ for the mass ratio of the ions
to electrons. This makes it easier to resolve small electron scales. In order
to lessen the separation between the field and particle timescales, we set the
ratio of the speed of light to the Alfv\'en speed, $c/c_A$, to $25$ (in the
heliosheath a more realistic value is near $6000$). Reconnection is insensitive
to the value of $c/c_A$. We start with a half-thickness for the current sheet
$w_0=0.5d_i$, so that collisionless reconnection can begin from particle noise.
The temperature in the Harris sheet is $0.25T_0$ for both ions and electrons,
and there is no guide field. The largest $\beta$ we simulated was $4.8$ since
the electron thermal velocity $v_{\text{the}} \approx 0.7c$. Larger $\beta$ would
begin to have significant unphysical relativistic effects, due to our lowered
ratio of $c/c_A$.

The simulation does not precisely describe the heliosheath but illustrates
important physics that should be found there. The ion pressure in the
heliosheath is much larger than the electron pressure, and thus $\beta_e$ is
actually quite moderate. However, we will show that at realistic mass ratios we
still expect elongated islands. The large $\beta$, mostly due to ion pressure,
is however significant because it allows the elongation to persist. We do not
have a separate population of PUIs, and the magnetic field
configuration is a Harris sheet rather than the rotated fields (where $|B|$ is
constant in a cut through the current sheet) found in the heliosheath. Future
simulations tailored to the specific parameters of the heliosheath may help our
understanding of this phenomenon.

\section{Results}

\begin{figure}
  \noindent\includegraphics[width=3.0in]{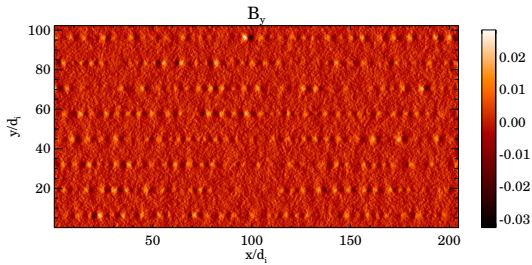}
  \caption{\label{by}
  Normal component of the magnetic field, $B_y$, for
  $\beta = 2$ at $t= 15\Omega_{\text{\text{ci}}}^{-1}$. The bipolar signatures in
  the current sheet indicate the presence of $x$-lines arising from the
  collisionless tearing mode.}
\end{figure}

The early development of a run with $\beta=2$ is shown in
Figure\,\ref{by}. Not surprisingly a wave mode with $k_xw_0 \approx 0.5$
clearly emerges. The finite $\beta$ background plasma does not have a
strong effect on the wavelength of linear tearing. During this time,
within the current sheets, an anisotropy in the electron pressure
begins to develop with $P_{e\parallel} > P_{e\perp}$. The electrons
moving at the thermal velocity are able to bounce between the two ends of the
islands which have lengths of $\sim 6 d_i$. Comparisons with runs at different
$\beta$ show that at this early time, the lengths of the islands appear to be
insensitive to $\beta$. 

There are two important timescales controlling the dynamics: the time
it takes for the ions to accelerate to Alfv\'enic outflow speeds and
the time it takes for a significant electron pressure anisotropy to develop.  If
electrons bounce several times between the two ends of a contracting
island, an anisotropy develops which approaches the firehose
instability boundary. This is because the bouncing electrons gain energy in the
parallel direction. The time for an electron to bounce off the edge of
an island and then return to its original position is thus a measure
of the time for significant anisotropy to develop.

The tearing instability is driven by the tension in the newly
reconnected magnetic fields. Since anisotropies cause a weakening of
the magnetic tension, the tearing mode can be suppressed by strong
anisotropy within an island.

Reconnecting magnetic field lines, by relaxing their tension, accelerate ions
up to Alfv\'enic speeds. If several bounces occur during the time required for
ions to be accelerated up to the Alfv\'enic outflow speed from the $x$-line, the
developing anisotropy slows the ion outflow and essentially stops the growth of
the tearing mode. However, since the bounce time is proportional to the length
of the islands, the growth of sufficiently long wavelength tearing modes can
continue.

\begin{figure}
  \noindent\includegraphics[width=3.0in]{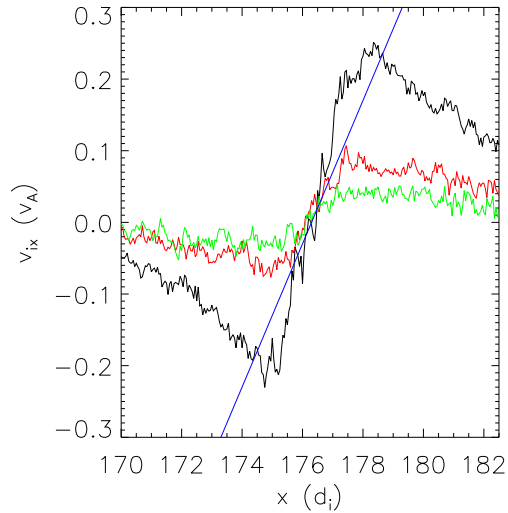}
  \caption{\label{growthtime}
    Ion outflow velocity, $v_{ix}$, vs. the position
    $x$, for $t=$ $25\Omega_{\text{ci}}^{-1}$ (green), $30\Omega_{\text{ci}}^{-1}$
    (red), and $35\Omega_{\text{ci}}^{-1}$ (black) for the $\beta=2$ run. To
    reduce noise we do a five-point smoothing of $v_{ix}$ in both the
    $\mathbf{\hat{x}}$ and $\mathbf{\hat{y}}$ directions. The blue
    curve is a line of slope $0.1$, which corresponds to a convective
    growth time $t_a$ of around $10\Omega_{\text{ci}}^{-1}$.}
\end{figure}

Near an $x$-line adjacent to a growing island, the outflow velocity of
ions, to first approximation,  linearly
increases with distance (see Figure\,\ref{growthtime}) as 
\begin{equation}
  v_{ix} = \frac{1}{t_a} x \text{.}
\end{equation} 
At this point $t_a$ is just defined as the inverse slope of the relationship
between $v_{ix}$ and $x$. Since according to Figure\,\ref{growthtime} the slope
is constant, this implies that ions accelerate away from the $x$-line in an
exponential fashion. By integrating, the time, $t$, for an ion to accelerate
from the initial position, $x_0$, to the final position, $x_f$, can be
obtained.
\begin{equation}
  t = \int_{x_0}^{x_f} \frac{dx}{v_{ix}} = \int_{x_0}^{x_f}
  \frac{t_a dx}{x} = t_a \ln{\frac{x_f}{x_0}} \text{,}
\end{equation}
and thus
\begin{equation}
  x_f = x_0 e^{t/t_a} \text{.}
\end{equation}
An approximate measure for the characteristic timescale for acceleration away
from the $x$-line up to the Alfv\'en speed is $t \approx t_a$, the acceleration
time.  As seen in Figure\,\ref{growthtime}, the acceleration time at $t = 30
\Omega_{\text{ci}}^{-1}$ is of order $\sim 10\Omega_{\text{ci}}^{-1}$. This acceleration time
is approximate, can vary by a factor of as much as two, and appears to be
insensitive to $\beta$.  

The bounce time can be estimated based on the thermal velocity of the
electrons, $v_{\text{the}}$, and the length of the island, $L$:  
\begin{equation}
  \label{btime} 
  t_{b} = \frac{L}{v_{\text{the}}} =
  \frac{L}{v_A}\sqrt{\frac{1}{\beta_e}\frac{m_e}{m_i}} \text{,}
\end{equation}
where $\beta_e$ is the $\beta$ determined solely from the plasma pressure
derived from the electrons. 

Equating the empirical acceleration time $t_a=10\Omega_{\text{ci}}^{-1}$, and the bounce time $t_b$,
Equation\,\eqref{btime}, a critical island length can be found:
\begin{equation}
  L_{crit} \approx
  10d_i\sqrt{\beta_e\frac{m_i}{m_e}} \text{.}
\end{equation}
For islands with $L < L_{crit}$, the anisotropy will stop the tearing
instability. Islands smaller than $L_{crit}$ can still form, but they
quickly saturate. A similar saturation was found in
\cite{Karimabadi05}. However, in their simulations the
size of the computational domain was $12.6\rho_i$ and $L_{crit}=100\rho_i$, where
$\rho_i$ is the ion Larmor radius. Thus, the development of long
wavelength islands was not observed.

\begin{figure}
  \noindent\includegraphics[width=3.0in]{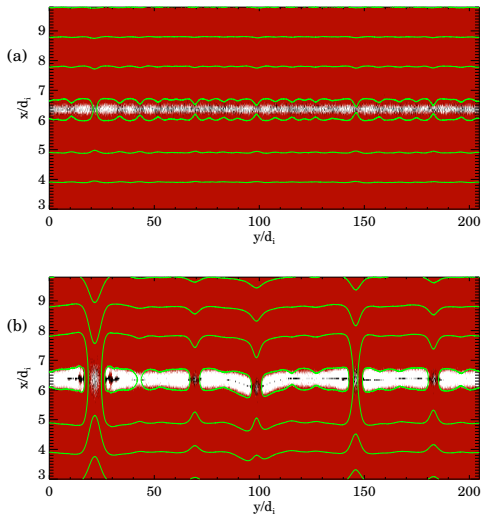}
  \caption{\label{anisotropyfound}
  Regions where the plasma anisotropy causes instability. White
  regions are unstable to the firehose, black regions are unstable to
  the mirror-mode, and red are stable. The green lines are magnetic
  field lines. This plot shows one current layer taken from the
  $\beta=2$ run at (a) $t=25\Omega_{\text{ci}}^{-1}$ and (b)
  $t=40\Omega_{\text{ci}}^{-1}$. The aspect ratio is distorted to make the
  islands more visible.}
\end{figure}

For the case of $\beta = 2$ ($\beta_e=1$) and $L \approx 6d_i$, $t_b
\approx 1.2\Omega_{\text{ci}}^{-1}$. This time is much less than the
acceleration time, so there is enough time for a significant
anisotropy to develop before a significant $x$-line is established. This
anisotropy can be seen in Figure\,\ref{anisotropyfound}(a),
which shows the regions from the $\beta=2$ run that are unstable to
the firehose instability. The unstable regions occur inside the
islands and stop further growth of the short wavelength tearing modes.
The islands that continue to grow correspond to longer
wavelength, with $L \approx 40d_i$ and $t_b \approx
8\Omega_{\text{ci}}^{-1} \approx t_a$. Thus, the anisotropy develops slowly enough
for reconnection to develop. This can be seen in
Figure\,\ref{anisotropyfound}(b).

\begin{figure}
  \noindent\includegraphics[width=3.0in]{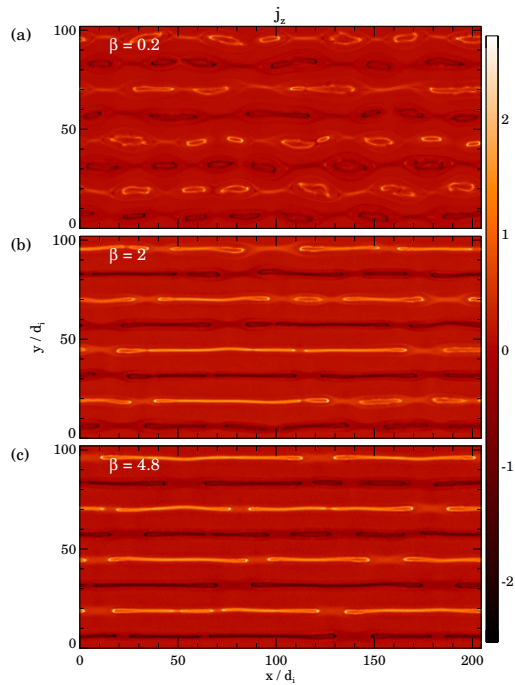}
  \caption{\label{jzbeta}
  Out-of-plane current, $j_z$, at $t = 51
  \Omega_{\text{ci}}^{-1}$ for $\beta$ of (a) 0.2, (b) 2, and (c) 4.8.}
\end{figure}

\begin{figure}
  \noindent\includegraphics[width=3.0in]{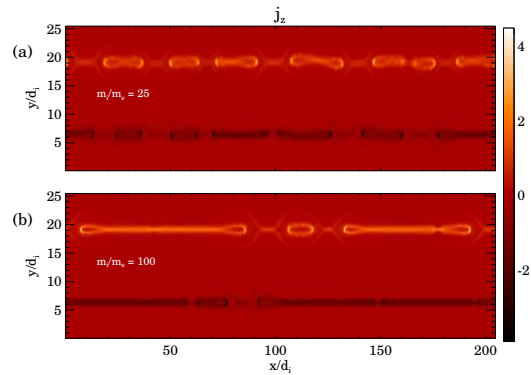}
  \caption{\label{islandme} 
  Out-of-plane current, $j_z$, for $\beta = 0.2$ at (a) $t = 40
  \Omega_{\text{ci}}^{-1}$ for $m_i/m_e = 25$ and (b) $t = 60
  \Omega_{\text{ci}}^{-1}$ for $m_i/m_e = 100$. The aspect ratio is distorted
  to make the islands more visible.}
\end{figure}

As can be seen in Figure\,\ref{jzbeta}, by $t=51\Omega_{\text{ci}}^{-1}$, $\beta$
has a significant influence on the structure of islands. The islands
for $\beta = 0.2$ have much shorter wavelength than for $\beta = 2$
and $4.8$. In other words, there are more locations where reconnection
proceeds in the case of low $\beta$. This phenomenon is expected based
on the previous analysis, $L_{crit} \propto \sqrt{\beta_e}$. Since $L_{crit}$
is proportional to the square root of the mass ratio $\sqrt{m_i/m_e}$, we
expect to find much longer islands in the real mass ratio limit. To test this,
we perform a $\beta = 0.2$ simulation with $m_i/m_e = 100$. In this case, we
reduce the $y$-domain by a factor of four with respect to Figure\,\ref{jzbeta}(a),
examining only two current sheets. We double the resolution in order to resolve
the small electron scales and reduce the ratio of the speed of light to the
Alfv\'en speed to $15$. There is a clear dependence on $m_i/m_e$ shown in
Figure\,\ref{islandme}, where we compare the bottom two current sheets of
Figure\,\ref{jzbeta}(a) to the new simulation.  We find the islands to be
significantly longer, confirming our prediction.  Since $m_i/m_e \gg 100$ in
the heliosphere, long islands are almost always expected, unless $\beta_e$ is
very small.

\begin{figure}
  \noindent\includegraphics[width=3.0in, trim = 0 3cm 0 0,clip]{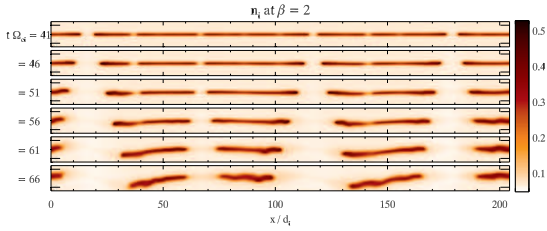}
  \caption{\label{islandcontraction}
  Ion density, $n_i$, for $\beta = 2$, along one current sheet
  between $t=41$ and $66\Omega_{\text{ci}}^{-1}$.}
\end{figure}

As elongated islands grow at high $\beta$, anisotropies within them
also develop even though the anisotropies do not suppress island
growth. The anisotropy surpasses the firehose
condition in the center of the islands. These anisotropies are likely
caused by the Fermi mechanism \citep{Drake06a}. The dynamics of this acceleration
mechanism will be discussed in a future paper. The contraction of
islands can be seen in Figure\,\ref{islandcontraction}. The higher
density regions inside of the islands move inward at Alfv\'enic
speeds. At around $t = 61-66\Omega_{\text{ci}}^{-1}$ the islands begin to
kink, which indicates the onset of an anisotropy instability.

The short-wavelength mode is caused by the temperature anisotropy due to the
outflow from the $x$-line streaming through the plasma entering the exhaust
across the separatrix and the Fermi acceleration of electrons bouncing in the
island.  Based on the similarities in growth rate and other signatures that will
be discussed in a future paper, this mode appears to be associated with the
Weibel instability.

\begin{figure}
  \noindent\includegraphics[width=3.0in]{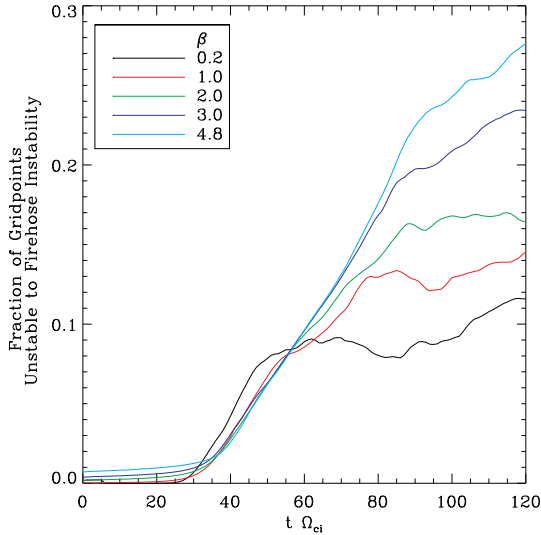}
  \caption{\label{FHIextent}
  Fraction of grid points unstable to the firehose
  instability vs. time for $\beta = 0.2, 1, 2, 3,$ and $4.8$.}
\end{figure}

The anisotropies that develop during the reconnection simulation do not grow
without bound. In Figure\,\ref{FHIextent}, we plot the fraction of grid points
that are unstable to the firehose instability. As time advances and the
anisotropies begin to form, the number of grid points unstable to the firehose
instability increases. However, at $t \sim 80\Omega_{\text{ci}}^{-1}$, the number of
unstable grid points begins to saturate. Since it takes place soon after the
onset of the kinking of the islands, the saturation is likely because the
anisotropy is reduced via scattering by the Weibel and firehose instabilities.
Additionally the saturation occurs soon after the unreconnected flux is
exhausted.  By $60 \Omega_{\text{ci}}^{-1}$ the islands have grown enough so that the
islands on adjacent current sheets begin to interact. This is an additional
reason for the saturation of the firehose unstable area: there is no more space
into which the firehose unstable islands can expand. 

\begin{figure}
  \noindent\includegraphics[width=3.0in]{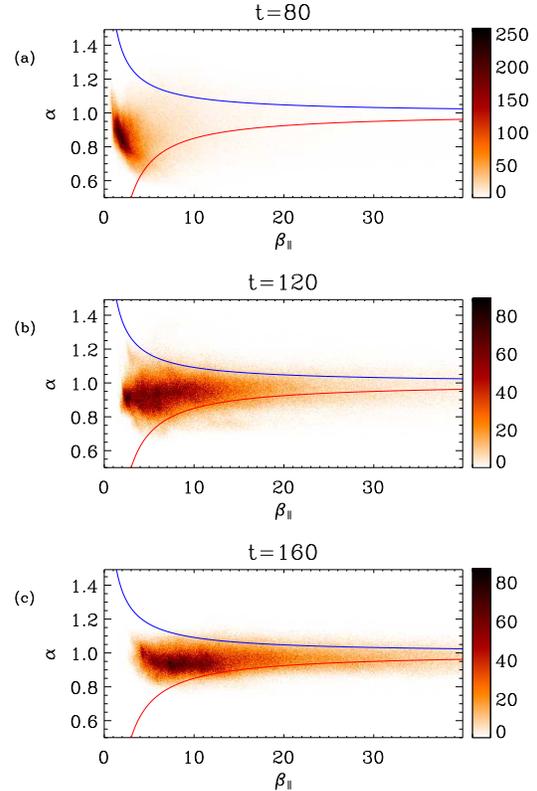}
  \caption{\label{firedist}
  Two-dimensional histogram of anisotropy ($\alpha = P_\perp/P_\parallel$) vs.
  $\beta_\parallel$ for $\beta = 2$ at times from top to bottom $t =
  80, 120,$ and $160\Omega_{\text{ci}}^{-1}$. The blue line represents the
  marginal condition for mirror-mode instability. Points above this
  curve are unstable. The red line represents the firehose marginal
  stability condition. Points below this curve are unstable. The color
  bar represents the number of points with a particular $\alpha$ and
  $\beta_{\parallel}$.}
\end{figure}

At late time, the anisotropy of the system is confined within the
boundaries of the marginal firehose (Equation\,\eqref{firehose}) and
mirror-mode instabilities (Equation\,\eqref{mirrormode}) in a manner similar to
that seen in observations of the solar wind \citep{Hellinger06, Bale09}
and in earlier low-$\beta$ current sheet simulations \citep{Drake10}.
Figure\,\ref{firedist} shows the data for our system in the space of
($\alpha$,$\beta_\parallel$) where $\alpha = P_\perp/P_\parallel$.
This plot is generated by calculating the anisotropy $\alpha$ and the
$\beta_{\parallel}$ for each grid point. The plot is a two-dimensional
histogram of grid points in ($\alpha$,$\beta_{\parallel}$) space, where
$\beta_{\parallel}$ is calculated based on $P_\parallel$. The parallel
and perpendicular pressures are calculated by taking the diagonal
components of the pressure tensor after rotating into the frame of the
local magnetic field, such that the two perpendicular components are
equal. We look at the distribution at $t = 80, 120,$ and
$160\Omega_{\text{ci}}^{-1}$. At early times the anisotropies have not yet
fully developed and the plasma still occupies a small region in
($\alpha$,$\beta_{\parallel}$) space. By $t=120\Omega_{\text{ci}}^{-1}$ the
anisotropy has reached the two stability boundaries, and continues to
be confined between these two boundaries at $t=160\Omega_{\text{ci}}^{-1}$,
even as the average $\beta$ increases. The anisotropy reaches the
stability boundaries at a time after the short wavelength Weibel modes
have dissipated. Since at this point there are no longer large regions
with essentially zero magnetic fields, the firehose and mirror-mode
instabilities are what determine the boundaries of the temperature
anisotropies. There are no clear signatures of the classical mirror-mode
instability at this time. The firehose and mirror-mode instabilities may
be hard to distinguish among the turbulent interacting magnetic islands, or the
islands may just stop generating anisotropy as they approach the instability
boundaries.

\begin{figure}
  \noindent\includegraphics[width=3.0in]{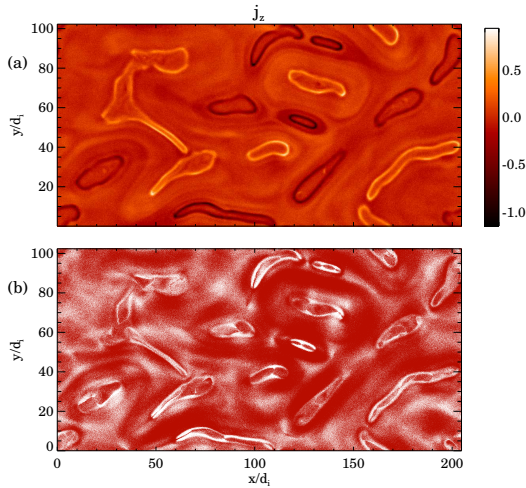}
  \caption{\label{jzanisotropy}
  Out-of-plane current, $j_z$(a), and stability(b)
  at $t= 120\Omega_{\text{ci}}^{-1}$ for the $\beta=4.8$ case.
  White regions in part (b) are unstable to the firehose, and red are stable.}
\end{figure}

The islands maintain an elongated form for the simulation shown in
Figure\,\ref{jzbeta}(c) clear until $t=120\Omega_{\text{ci}}^{-1}$, the latest time
simulated for $\beta=4.8$. This is shown in Figure\,\ref{jzanisotropy}(a) showing
the out-of-plane current for $t=120\Omega_{\text{ci}}^{-1}$. Since the edges of the
islands are pushing against the firehose instability, the tension force in the
magnetic fields is eliminated. This can be seen in Figure\,\ref{jzanisotropy}(b)
which shows the regions that are unstable to the firehose instability.

\section{Conclusions}

The magnetic islands that reach a significant amplitude are much
more elongated at high $\beta_e$ than at low $\beta_e$. These
elongated islands should be found even for moderate values of
$\beta_e$ at realistic mass ratios. Island elongation is caused by the
suppression of the shorter wavelength tearing modes by pressure
anisotropies ($P_\parallel > P_\perp$) that develop due to the Fermi
acceleration of electrons. Later in time the plasma develops
pressure anisotropies of both ions and electrons that are
limited by the firehose and Weibel instabilities. A Weibel mode
develops that kinks the magnetic field lines. In the regime with a
real mass ratio we would expect even longer islands to form, where
multiple wavelengths of the firehose instability could develop. At late
time the fraction of points unstable to the firehose instability
saturates, and the anisotropy is confined between the mirror-mode and firehose
instability boundaries. The long islands persist due to the low requirement of
anisotropy to reach the marginal firehose condition at high $\beta$. For even
small anisotropies the tension in the magnetic fields is removed.

When encountering magnetic islands in the heliosheath, we predict the formation
of similar extended, sausage-shaped islands rather than the more round islands
found in low-$\beta$ simulations \citep{Drake10}. The cores of these islands
should also be at the marginal firehose condition, so the magnetic tension that
drives them to become round vanishes. We would thus expect these sausage
shapes to persist long after the islands have ceased growing, and thus could be
found even in regions where reconnection is no longer occurring.

\begin{figure}
  \noindent\includegraphics[width=3.0in]{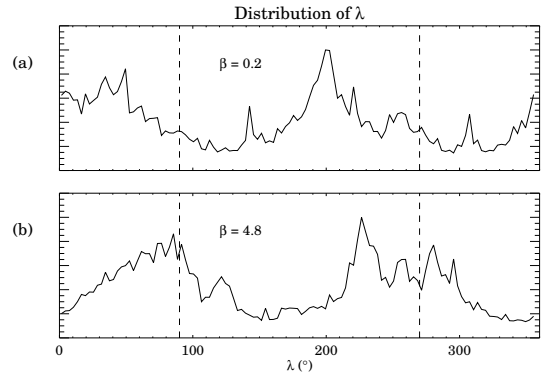}
  \caption{\label{lambdacompare}
  Distribution of $\lambda$ at $t= 110\Omega_{\text{ci}}^{-1}$ for the (a)
  $\beta=0.2$ case and (b) $\beta=4.8$ case. The dotted lines are at $\lambda =
  90^\circ$ and $270^\circ$ where we expect to find peaks in the distribution.}
\end{figure}

These elongated islands exhibit signatures that can be seen in \emph{Voyager}
data. In particular, \emph{Voyager} measures all three components of the magnetic
field. Of particular interest for the explorations of islands that grow in the
ecliptic plane is the angle $\lambda =\tan^{-1}\left(B_T/B_R\right)$, where
$B_T$ and $B_R$ are the azimuthal and radial magnetic fields, respectively.
$\lambda = 90^\circ$ and $270^\circ$ correspond to the azimuthal unreconnected
sectored heliosheath magnetic fields. Deviation of $\lambda$ from $90^\circ$
and $270^\circ$ indicates some process is distorting the sectored field.
\emph{Voyager} data show the distribution of $\lambda$ is peaked in the two azimuthal
directions, $\lambda = 90^\circ$ and $270^\circ$ \citep{Opher11}. These peaks
are significantly broader in the heliosheath than upstream, indicating that
reconnection or another mechanism is disturbing the heliosheath field. The
observed \emph{Voyager} distribution of $\lambda$ is consistent with that found in
high-$\beta$ simulations\citep{Opher11}. Since the islands are elongated, the
magnetic fields tend to remain primarily in the azimuthal direction even well
after the islands begin to interact with each other. Round islands, such as
would be expected from an MHD model or a low $\beta$ kinetic model, are not
consistent with observations since they produce much broader $\lambda$
distributions. Thus, MHD reconnection \citep{Lazarian09} in the heliosheath
seems to be ruled out. Shown in Figure\,\ref{lambdacompare} is the distribution
of $\lambda$ from the simulations at $\beta=0.2$ (Figure\,\ref{jzbeta}(a)), and
$\beta=4.8$ (Figure\,\ref{jzbeta}(c)) at $t=110 \Omega_{\text{ci}}^{-1}$. The high
$\beta$ simulation which has elongated islands retains the two peaks at
$\lambda=90^\circ$ and $\lambda=270^\circ$. The long islands have a larger
magnetic field in the azimuthal direction than the radial, resulting in peaks
in the $\lambda$ distribution, but the shorter islands become round having a
magnetic field with similar strength in both directions, resulting in a broad
distribution in $\lambda$. The loss of tension in a finite $\beta$ plasma
prevents the complete release of magnetic energy that would be expected in an
MHD model. A complete understanding of the $\beta$ dependence of magnetic
islands is essential in order to obtain reliable signatures that can be
compared with \emph{Voyager} data.

In this work there was no out-of-plane guide magnetic field. In the
heliospheric current sheet, the magnetic field rotates from one direction to
the other keeping a constant magnitude rather than passing through zero
\citep{Smith01}. A guide field would cause the center of the islands to have a
much lower $\beta$ since the magnetic field does not go to zero. Because of
this magnetic field, we would not expect the Weibel instability to develop. In
real systems there is frequently a guide field, so this would be worth further
investigation.

In astrophysical accretion disks, reconnection plays a role in determining the
saturation of the magnetorotational instability (MRI) \citep{Sano04}. The
saturation of MRI is strongly dependent on the dissipation of the magnetic
field due to reconnection. For the high $\beta$ in accretion disks, suppression
of the most strongly growing small islands may significantly impact the
saturation of the MRI. Since $\beta$ is typically larger than $100$ in these
structures, the only surviving islands would be so long that it is likely
that much of the magnetic free energy would not be dissipated. Further, since
the MRI requires magnetic tension, the absence of tension could limit the
development of the instability.  \cite{Sharma06} perform a simulation showing
an enhancement of the growth of MRI due to anisotropies with $P_\perp >
P_\parallel$ , which enhances the magnetic tension, caused by $\mu$
conservation as a magnetic field develops. They do not capture the physics of
reconnection and Fermi acceleration in magnetic islands that would generate
anisotropies with $P_\perp < P_\parallel$, which removes magnetic tension.
These two competing sources of anisotropy, both affect the tension and thus the
growth of the MRI. The relative importance of these mechanisms needs to be
explored. 

Reconnection at high $\beta$, although relatively rare in the terrestrial
magnetosphere, is also found in the magnetosphere of Saturn \citep{Masters11}.
Magnetic islands were discovered in a region where $\beta$ is larger than 10.
The $\beta$ dependence of the growth of finite-sized magnetic islands may lead
to a better understanding of these findings.

Since the development of elongated islands requires only moderate $\beta$, we
expect to see the development of longer islands than expected in lower $\beta$
systems such as the magnetosphere. In contrast to \cite{Karimabadi05}, these
longer islands can grow to a large enough size to play a role in magnetospheric
dynamics. Since the $\beta$ of the magnetosphere is not exceptionally large it is
unlikely that the persisting anisotropy is enough to keep the islands from
eventually becoming round.

\smallskip

The computations were performed at the National Energy Research Scientific
Computing Center.

                                                                                                                                         1,1           Top

\end{document}